# Structural identification of cubic iron-oxide nanocrystal mixtures:
## X-ray powder diffraction versus quasi-kinematic transmission electron microscopy

Peter Moeck

Department of Physics, Portland State University, P.O. Box 751, Portland, OR 97207-0751
& Oregon Nanoscience and Microtechnologies Institute, http://www.onami.us

Two novel (and proprietary) strategies for the structural identification of a nanocrystal from either a single high-resolution (HR) transmission electron microscopy (TEM) image or a single precession electron diffraction pattern are proposed and their advantages discussed in comparison to structural fingerprinting from powder X-ray diffraction patterns. Simulations for cubic magnetite and maghemite nanocrystals are used as examples. This is an expanded and updated version of a conference paper that has been published in Suppl. Proc. of TMS 2008, 137th Annual Meeting & Exhibition, Volume 1, Materials Processing and Properties, pp. 25-32.

## Introduction

Nanocrystals possess size [1] and morphology [2] dependent properties that are frequently superior to those of the same materials in their bulk form. Any future large-scale commercial "nanocrystal powder-based industry" will need to be supported by structural assessment methods [3]. The quite ubiquitous method of identifying crystal structures is (Cu-tube based) powder X-ray diffraction (XRD) [4], e.g., Fig. 1.

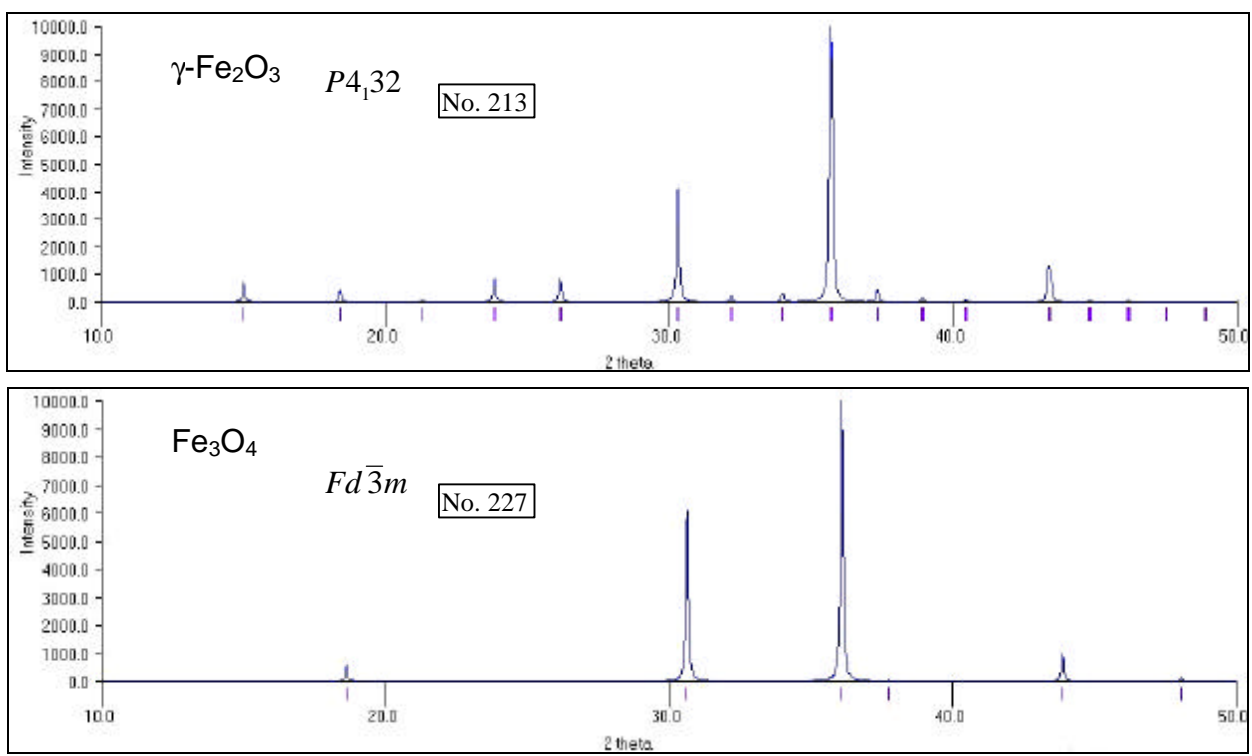

**Figure 1:** Calculated powder X-ray diffraction patterns for micrometer-sized cubic maghemite, γ-$Fe_2O_3$ (top), and magnetite, $Fe_3O_4$ (bottom), out to the 133 reflections. The space group symbols and their numbers are also given.

That method works best for micrometer-sized crystals and becomes less useful to useless for crystals in the nanometer-size range due to peak broadening and (isotropic or anisotropic) peak shifting [5,6]. XRD patterns of nanocrystals are also made significantly less characteristic by surface relaxation [7]. Two novel (and proprietary [8]) strategies for the structural identification of nanocrystals in the TEM are, therefore, proposed. Both of these strategies work best for nanocrystal thicknesses at which the scattering of fast electrons can be considered as essentially kinematic or quasi-kinematic. This thickness range is for HRTEM imaging 1 to about 10 nm and for PEDs 10 to 50 nm. In the dynamic scattering limit, these methods become analogous to the well-known structural identification methods for single crystals in the TEM that only use information on the projected reciprocal lattice geometry. For a recent review of those methods and more information on the two novel strategies, see ref. [9].

Because the crystal structures of cubic maghemite and magnetite possess almost the same lattice constant and a "similar" atomic arrangement (i. e. nearly cubic densest packings of oxygen with differences in the iron occupancies of the intersites), their powder XRD patterns are very similar, Fig. 1. Allowing for peak broadening, peak shifting, and surface relaxation, nanometer-sized crystals of these two iron-oxide minerals can hardly be told apart and their mixtures can not be quantified by XRD.

For mixtures of nanometer-size crystals, there is, however, for each single nanocrystal atomic-level structural information in its HRTEM image or precession* electron diffractogram. This information can be advantageously employed for structural identification [9-14]. The atomic-level structural information is in the case of HRTEM images (after crystallographic image processing [15]) structure factor amplitudes and phase angles out to the point resolution of the microscope, e.g. at least out to 5 $nm^{-1}$ for dedicated (non-aberration corrected) HRTEMs. In the case of precession electron diffractograms, this atomic-level structural information is structure factor amplitudes out to approximately twice as far in reciprocal space. As precession electron diffraction avoids crystal orientations that result in the simultaneous excitation of more than one strong diffracted beam (as much as this is possible), quasi-kinematic reflection intensities [16,17] are obtained for nanocrystals with thicknesses up to approximately 50 nm. Simultaneously present reflections in higher order Laue zones and systematic absences in both the higher and the zero order Laue zones allow frequently for an unambiguous determination of the space group [18]. Visual comparisons of kinematical electron diffraction simulations with experimental precession electron diffractograms have, therefore, been used for structure verifications [19]. Commercial IBM-PC based computer programs are available for such simulations**.

The more systematic approach of extracting structural information from HRTEM images or precession electron diffractograms of unknowns and comparing it directly to structural information that is contained in a crystallographic database allows for the identification of the unknowns [8]. For the extraction of structural information from both kinds of experimental data sources, one can advantageously employ commercial, IBM-PC based electron crystallography software***. As an alternative to the commercial databases, one may use their on-line demonstration versions [20,21], a free structure data request service for organics [22], or open-access databases [23-26], which provide together some 100,000 crystal structure data sets.

This paper illustrates that for nanocrystals which scatter fast electrons kinematically or quasi- kinematically much more structural information can be extracted from either HRTEM images or electron precession diffractograms than is accessible from powder XRD. Simulations for nanocrystals of cubic maghemite, γ-$Fe_2O_3$, and magnetite, $Fe_3O_4$, are used as examples. There are also tetragonal maghemites with similar stoichiometries and variations in the occupancy of the iron intersites, which are not considered here.

### Structural information from HRTEM images or precession electron diffractograms

Table I lists theoretical structure factor amplitudes and phase angles that are for cubic maghemite and magnetite nanocrystals. Their experimental counterparts can be extracted from HRTEM images that were recorded at a microscope with 0.19 nm point resolution. Figure 2 shows a so-called "lattice-fringe fingerprint plot" for magnetite for the same point resolution. This plot was calculated over the Internet (on the fly) from data of the (mainly inorganic and educational) subset of the Crystallography Open Database [26], Fig. 3. We call these plots "lattice-fringe fingerprint plots" because the idea to plot two reciprocal spacings and their acute intersecting angle, (i.e. 3 independent entities), into a two-dimensional (2D) plot originated in connection with Fourier transforms of HRTEM images that showed crossing lattice fringes [14].

As far as the projected reciprocal lattice geometry of the zero order Laue zone is concerned, there is no conceptional difference between lattice-fringe fingerprint plots that originated from Fourier transformed HRTEM images and their counterparts that originated from precession electron diffraction data. Lattice-fringe fingerprint plots from precession electron diffraction data do, however, extend much further into reciprocal space. Dedicated "higher order Laue zone lattice-fringe fingerprint plots" can also be constructed for precession electron diffraction data, but are not discussed further here.

**Table I**: (below, left column). Estimated**** theoretical structure factor amplitudes (structure factor magnitude |F| in nm) and phase angles ($\alpha$ in degree) for cubic maghemite, $\gamma$-$Fe_2O_3$, and magnetite, $Fe_3O_4$, which are extractable from Fourier transformed HRTEM images that were taken at a microscope with 0.19 nm point resolution.

| {hkl} | $\gamma$-$Fe_2O_3$ \|F\| | $\gamma$-$Fe_2O_3$ $\alpha$ | $Fe_3O_4$ \|F\| | $Fe_3O_4$ $\alpha$ |
|---|---|---|---|---|
| **011** | 0.78 | 90 | - | - |
| **111** | 0.55 | 135 | 1.55 | 0 |
| **012** | 0.90 | 90 | - | - |
| **112** | 0.60 | 0 | - | - |
| **022** | 3.25 | 0 | 3.29 | 180 |
| **013** | 0.50 | 270 | - | - |
| **113** | 4.41 | 45 | 4.85 | 180 |
| **222** | 0.15 | 90 | 1.11 | 0 |
| **023** | 0.63 | 0 | - | - |
| **123** | 0.43 | 180 | - | - |
| **004** | 5.65 | 180 | 6.47 | 0 |
| **033** | 0.38 | 270 | - | - |
| **114** | 0.38 | 270 | - | - |
| **133** | 0.28 | 135 | 0.37 | 180 |

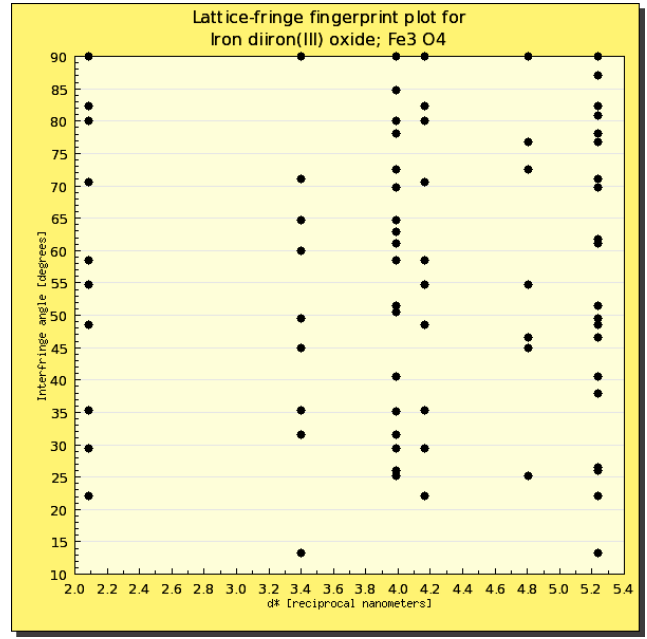

**Figure 2**: (above, right column). Lattice-fringe fingerprint plot of magnetite for a HRTEM with 0.19 nm point resolution.

Since it is a relatively new concept [8-14], lattice-fringe fingerprint plots need to be explained in some more detail. The so-called "interfringe angle", i.e. the acute angle under which lattice fringes intersect in HRTEM images, is plotted in such plots against the reciprocal lattice vector magnitude. While there are two data points in lattice-fringe fingerprint plots for crossed fringes with different spacings, the crossing of two symmetrically related fringes results in just one data point (because the latter possesses by symmetry the same spacing). These plots may extend in reciprocal space out to either the point or the instrumental resolution of the microscope. All of the resolvable lattice fringes up to this resolution will be included, for a certain crystal structure, into these plots. If derived from precession electron diffraction data, the counterpart to a lattice-fringe fingerprint plot will extend in reciprocal space out to the diffraction limit of the structure.

An initial search in a database of theoretical lattice-fringe fingerprints that is only based on the two-dimensional (2D) positions of lattice-fringe data points, Fig. 2, may result in several candidate structures. In a follow-up step, the search can be made more discriminatory by trying to match crystallographic indices to the 2D positions. Because one will always image along one zone axis, all of the indices of the reflections must be consistent with a certain family of zone axes. As far as the lattice-fringe fingerprint plots are concerned, this follow-up search is equivalent to assigning crystallographic indices to the 2D data points. This can be done on the basis of Weiss' zone law for the zero-order Laue zone (and also for higher-order Laue zones in case of precession electron diffraction data). Each (vertical) column of data points in a lattice-fringe fingerprint plot corresponds to one family of reflections (net-planes). Discrete points on a second x-axis in a lattice-fringe fingerprint plot could, thus, be labeled with the respective Miller indices, {h,k,l}, of a family of reflections. Each (horizontal) row of data points in a plot such as Fig. 2, on the other hand, belongs to a family of zone axes. Discrete points on a second y-axis of such a plot could, thus, be labeled with the respective Miller indices, <u,v,w>, of a family of zone axes. The cross product of the Miller indices of two data points from two different columns (representing two different reciprocal spacings) that are also located within the same row (representing one interfringe angle) gives the zone axis symbol, <u,v,w> = {$h_1,k_1,l_1$} x {$h_2,k_2,l_2$}. While each family of reflections would only show up once on such a second x-axis, the same family of zone axis symbol could show up multiple times on such a second y-axis. Guided by the added Miller indices for columns and rows on such a lattice-fringe fingerprint plot, kinematically forbidden reflections can be identified. (Because all interfringe angles between identically indexed reflections are the same in the cubic system, space group information can be extracted straightforwardly from lattice-fringe fingerprint plots of cubic crystals even without indexing.) Higher orders of a family of net-planes {n·h,n·k,n·l}, possess an (n times) integral multiple of the spatial frequency of that family. Such higher orders of families are also easily spotted in a lattice-fringe fingerprint plot because their "columns" look identical. This is because the respective higher order net-planes will intersect other net-planes at precisely the same interfringe angles as those net-planes from a lower order. Within the error bars and especially in lattice-fringe fingerprint plots for a very high microscope resolution, it is possible that families of net-planes or zone axes coincide on the second x- or y-axes. More elaborate lattice-fringe fingerprint plots may contain in the third and forth dimension information on structure factor phases and amplitudes. Possibly in a fifth dimension, histograms of the probability of seeing crossed lattice fringes in an ensemble of nanocrystals may be added to both types of lattice-fringe fingerprints and may facilitate the structural fingerprinting of a multitude of nanocrystals. The equations for calculating such probabilities for an ensemble of randomly oriented nanocrystals are given in ref. [14]. Instead of employing higher dimensional spaces, one could also stick to 2D displays such as Fig. 2 and simply add to selected data points sets of numbers that represent additional information, e.g. structure factor phase angles and amplitudes with their respective error bars.

Similarly to the classical Hanawalt search strategy of powder X-ray diffraction databases [4], one can divide lattice-fringe fingerprint plots, such as the ones shown in Fig. 2, into 2D geometric data sectors of experimental condition-specific average precisions and accuracies and also allow for some overlap between the sectors. Larger reciprocal spacings and interfringe angles can be measured inherently more accurately and precisely than smaller reciprocal spacings and interfringe angles. The location of the respectively more precise and accurate data points will be in the upper right-hand corners of lattice-fringe fingerprint plots.

The accuracy and precision of the extracted structure factors will depend on how accurately the contrast transfer function of the objective lens can be determined at every point of interest by crystallographic image processing [15]. The accuracy of theoretical structure factors is not precisely known as it depends on the (not precisely known) accuracy of the atomic scattering factors. Nevertheless, the accuracy of theoretical structure factors is likely to be similar for all structure factors because each of them represents the scattering in a certain direction by all of the atoms in the unit cell.

## Comparison between TEM and XRD data for structural identification of nanocrystals

If one takes the peak position and peak height in a powder X-ray diffractogram as two pieces of information, there are just 12 such pieces for magnetite (including those from the very weak 222 peak next to the strong 400 peak, Fig. 1), that can be used for the structural identification of this mineral. In Fig. 2, there are, however, 74 data points for magnetite out to the family of {133} reciprocal lattice vectors. In addition, each of the families of lattice planes in Fig. 2 possesses both structure factor amplitude and a structure factor phase angle.

Due to the primitive cubic space group symmetry of maghemite, its lattice-fringe fingerprint plot counterpart to Fig. 2 contains about 5 times more data points for the same point resolution of the HRTEM. In addition, there is no restriction of the structure factor phase angles to either 0° or 180° for maghemite due to its space group being not centrosymmetric. Table 1 shows noticeable differences in the structure factor amplitudes and phase angles for these two "rather similar" cubic iron-oxide minerals. Magnetite and cubic maghemite nanocrystals as part of a mixture are, thus, reliably distinguished on the basis of HRTEM images, as experimentally demonstrated in refs. [10-12].

If the counterparts of lattice-fringe fingerprint plots are for either maghemite or magnetite constructed from precession electron data, there will be many more data points in the plots, as the resolution of such data is not restricted to the point or information limit resolution of the HRTEM. There will, however, be for each family of lattice planes only the (real) structure factor amplitude available for structural fingerprinting in the TEM, because the (complex) phase angle information will be lost in the process of recording the diffraction pattern.

Since the indices of the three strongest peaks in XRD patterns out to the 133 reflection, Fig. 1, are for cubic maghemite and magnetite identical, these two iron-oxides can, even for micrometer-sized crystals, not easily be distinguished by the classical Hanawalt [4] approach. Due to XRD peak broadening, peak shifting, and surface relaxation effects, both a distinction between these two minerals and quantification in case of a mixture of these two iron-oxides become quite impossible for nanocrystals.

**Figure 3:** Access screen to the open-access crystallographic databases that are housed at Portland State University's research servers, URL: nanocrystallography.research.pdx.edu/CIF-searchable, April 1$^{st}$, 2008.

## Summary of the kind of information that is obtainable from TEM for structural identification of nanocrystals

The structural information that can be extracted from an HRTEM image is the projected reciprocal lattice geometry, the plane symmetry group, and a few structure factor amplitudes and phases. Except for the structure factor phases, the same kind of information can be extracted from a precession electron diffractogram, but the information that can be used for structural fingerprinting is in this case not limited to the point or instrumental resolution of the TEM. In addition, symmetry information from higher-order Laue zones is frequently also extractable from precession electron diffractograms. Searching for these kinds of structural information in comprehensive databases and matching it with high figures of merit to that of candidate structures allows for highly discriminatory identifications of nanocrystals, even without additional chemical information as obtainable in analytical TEMs.

## Acknowledgments

This research was supported by congressional earmark funds to the Office of Naval Research and a grant from the Oregon Nanoscience and Microtechnologies Institute.

---

\* The precession electron diffraction technique is formally analogous to the well known X-ray precession diffraction technique (Buerger, M. J.: Contemporary Crystallography, McGraw-Hill, 1970, p. 149-185), but utilizes a precession movement of the electron beam around the microscope's optical axis rather than that of the specimen goniometer around a fixed-beam direction. The diffracted beams are de-scanned in such a manner that stationary spot diffraction patterns are obtained. The illuminating electron beam can be either parallel or focused. Precession electron add-ons to (newer and older) TEMs have been developed by Dr. Stavros Nicolopoulos and coworkers and can be purchased from the following Belgian company: NanoMEGAS SPRL, Boulevard Edmond Machterns No 79, Sint Jean Molenbeek, Brussels, B-1080, Belgium, Telephone: +34 649 810 619, http://www.nanomegas.com, info@nanomegas.com. **Portland State University's "Laboratory for Structural Fingerprinting and Electron Crystallography"** (run by Prof. Peter Moeck) **is the first demonstration site for NanoMEGAS SPRL in the Americas.** There is currently only one other commercial precession electron system from NanoMEGAS installed in the USA (at ExxonMobile Research & Engineering Co. Inc, Annandale, NJ), while there are already 28 installations in Europe and one in South Korea. Northwestern University (Evanston, IL) possesses a user-built precession electron system. Copies of that system have been installed at the University of Illinois at Urbana-Champaign, Arizona State University, and UOP LLC, a Honeywell company (Des Plaines, IL).

\*\* Simulations of precession electron diffractograms can be performed with the software Emap and Simulator by Dr. Peter Oleynikov from AnaliTEX, http://www.analitex.com. The simulated electron diffraction spot intensities can be toggled between proportionality to the square of the structure factor magnitude and proportionality to the structure factor magnitude. While the former simulates kinematic diffraction, the latter provides a rough estimate of dynamical diffraction effects without regards to the actual crystal thickness, orientation and shape. This AnaliTEX software can be demonstrated at Prof. P. Moeck's lab in Portland, OR.

\*\*\* Profs. Sven Hovmöller and Xiaodong Zou of the Swedish company Calidris, http://www.calidris-em.com, offer IBM-PC compatible software that supports the extraction of structural information from both HRTEM images and precession electron diffraction spot patterns. While structure factor phase angles and amplitudes can be extracted from HRTEM images with the program CRISP, their ELD software supports the extraction of electron diffraction intensities. Calidris' PhiDO program is part of ELD and allows for the identification of crystal phases from either one or two electron diffraction spot patters that were recorded with a known angular orientation difference. A space group determination program that utilizes structural information from both zero-order and higher-order Laue zones (for one or more crystal orientations) is also part of ELD. The Calidris electron crystallography software suite can be demonstrated at Prof. P. Moeck's lab in Portland, OR.

\*\*\*\* Estimated with AnaliTEX's Emap and Simulator (http://www.analitex.com), electron wavelength: 2.5 pm, corresponding to a TEM acceleration voltage of 200 kV.